\documentclass[aps,superscriptaddress,twocolumn,twoside,floatfix,pra,nofootinbib,a4paper]{revtex4-2}
\usepackage{enumerate,appendix}
\usepackage{amsmath, amsthm, amssymb,commath}
\usepackage{color,calc,graphicx}
\usepackage[usenames,dvipsnames,svgnames,table,cmyk,hyperref]{xcolor}
\usepackage[colorlinks]{hyperref}
\hypersetup{
	colorlinks = true,
	urlcolor = {blue},
	citecolor = {magenta},
	linkcolor= {blue}
}

\usepackage{graphicx}
\usepackage{amsmath}
\usepackage{latexsym}
\usepackage{bbm}

\newcommand{\ket}[1]{|#1\rangle}

\usepackage{physics}
\newcommand{\degree}{{}^{\circ}}
\usepackage[charter,cal=cmcal,sfscaled=false]{mathdesign}
\usepackage{booktabs}
\usepackage{multirow}
\usepackage{subfigure}
\usepackage{dcolumn}
\usepackage{mathrsfs}
\usepackage{diagbox}
 \usepackage{array}
 \usepackage{makecell}
\usepackage{threeparttable}

\usepackage{float}

\begin{document}

\title{Compression of Entanglement Improves Quantum Communication}

\author{Yu Guo}
\thanks{These authors contributed equally.}
\affiliation{CAS Key Laboratory of Quantum Information, University of Science and Technology of China, Hefei, 230026, China}
\affiliation{CAS Center For Excellence in Quantum Information and Quantum Physics, 	University of Science and Technology of China, Hefei, 230026, China}

\author{Tang Hao}
\thanks{These authors contributed equally.}
\affiliation{CAS Key Laboratory of Quantum Information, University of Science and Technology of China, Hefei, 230026, China}
\affiliation{CAS Center For Excellence in Quantum Information and Quantum Physics, 	University of Science and Technology of China, Hefei, 230026, China}

\author{Jef Pauwels}
\affiliation{Laboratoire d'Information Quantique, CP 225, Universit\'e libre de Bruxelles (ULB),\\ Av. F. D. Roosevelt 50, 1050 Bruxelles, Belgium}

\author{Emmanuel Zambrini Cruzeiro}
\affiliation{Instituto de Telecomunicações, Instituto Superior Técnico, 1049-001 Lisbon, Portugal}

\author{Xiao-Min Hu}
\affiliation{CAS Key Laboratory of Quantum Information, University of Science and Technology of China, Hefei, 230026, China}
\affiliation{CAS Center For Excellence in Quantum Information and Quantum Physics, 	University of Science and Technology of China, Hefei, 230026, China}

\author{Bi-Heng Liu}
\email{bhliu@ustc.edu.cn}
\affiliation{CAS Key Laboratory of Quantum Information, University of Science and Technology of China, Hefei, 230026, China}
\affiliation{CAS Center For Excellence in Quantum Information and Quantum Physics, 	University of Science and Technology of China, Hefei, 230026, China}
\affiliation{Hefei National Laboratory, University of Science and Technology of China, Hefei, 230088, China}

\author{Yun-Feng Huang}
\affiliation{CAS Key Laboratory of Quantum Information, University of Science and Technology of China, Hefei, 230026, China}
\affiliation{CAS Center For Excellence in Quantum Information and Quantum Physics, 	University of Science and Technology of China, Hefei, 230026, China}
\affiliation{Hefei National Laboratory, University of Science and Technology of China, Hefei, 230088, China}

\author{Chuan-Feng Li}
\affiliation{CAS Key Laboratory of Quantum Information, University of Science and Technology of China, Hefei, 230026, China}
\affiliation{CAS Center For Excellence in Quantum Information and Quantum Physics, 	University of Science and Technology of China, Hefei, 230026, China}
\affiliation{Hefei National Laboratory, University of Science and Technology of China, Hefei, 230088, China}

\author{Guang-Can Guo}
\affiliation{CAS Key Laboratory of Quantum Information, University of Science and Technology of China, Hefei, 230026, China}
\affiliation{CAS Center For Excellence in Quantum Information and Quantum Physics, 	University of Science and Technology of China, Hefei, 230026, China}
\affiliation{Hefei National Laboratory, University of Science and Technology of China, Hefei, 230088, China}

\author{Armin Tavakoli}
\email{armin.tavakoli@teorfys.lu.se}
\affiliation{Physics Department, Lund University, Box 118, 22100 Lund, Sweden}

\begin{abstract}

Shared entanglement can significantly amplify classical correlations between systems interacting over a limited quantum channel. A natural avenue is to use entanglement of the same dimension as the channel because this allows for unitary encodings, which preserve global coherence until a measurement is performed. Contrasting this, we here demonstrate a distributed task based on a qubit channel, for which irreversible encoding operations can outperform any possible coherence-preserving protocol. This corresponds to using high-dimensional entanglement and encoding information by compressing one of the subsystems into a qubit. Demonstrating this phenomenon requires the preparation of a four-dimensional maximally entangled state, the compression of two qubits into one and joint qubit-ququart entangled measurements, with all modules executed at near-optimal fidelity. We report on a proof-of-principle experiment that achieves the advantage by realizing separate systems in distinct and independently controlled paths of a single photon.  Our result demonstrates the relevance of high-dimensional entanglement and non-unitary operations for enhancing the communication capabilities of standard qubit transmissions.

\end{abstract}

\maketitle

\section{Introduction}
Two key resources for quantum communication are the transmission of a quantum state and shared entanglement. Each of these resources can create stronger-than-classical correlations over an information-theoretically limited channel (see e.g.~\cite{Tavakoli2015, Martinez2018, Ahrens2012, Hendrych2012, Prevedel2011, Muhammad2014, Buhrman2010}). However, the strongest quantum communication resource is the combination of both, namely entanglement-assisted quantum communication. This enables an increase in the classical capacity of quantum channels \cite{Bennett1999, Bennett2002}. Advantages are possible even when the channel is only used once, as in protocols like dense coding  \cite{Bennett1992}, where a sender can transmit two bits over a noise-free qubit channel by locally rotating and relaying one share of an entangled pair of qubits \cite{Mattle1996, Li2002, Barreiro2008, Williams2017, Hu2018}.

\begin{figure}[t!]
	\centering
	\includegraphics[width=0.9\columnwidth]{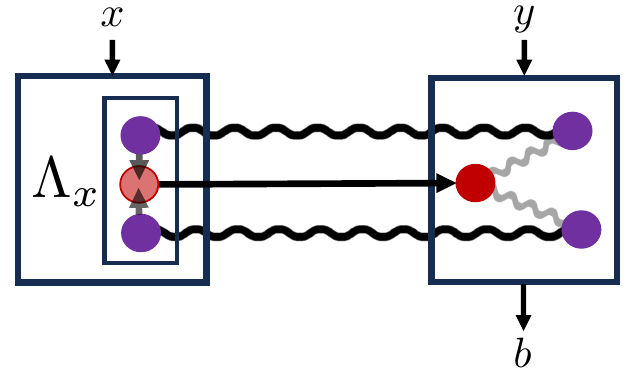}
	\caption{\emph{Qubit message assisted with high-dimensional entanglement.} The sender applies quantum channels $\Lambda_x$, compressing her share of a high-dimensional entangled state (purple) into a qubit (red) which is sent over the channel. The receiver measures the total received dimension-asymmetric entangled state by projecting onto an entangled basis. \label{fig:scenario}}
\end{figure}

High-dimensional entanglement cannot improve dense coding. This may be intuitive since using the same entanglement and channel dimension permits unitary encodings for the sender. Until the measurement is performed, the entanglement of the two shares is preserved. In contrast, high-dimensional entanglement requires irreversible encoding operations which compress the share into a qubit before it is transmitted. The compression can be seen as a unitary followed by binning part of the state, decohering the entanglement in the process and imposing an asymmetry between channel and source dimensions (see Figure~\ref{fig:scenario}). Interestingly, the intuition that coherent operations are optimal does not hold for more general tasks beyond sending and receiving a classical input \cite{Tavakoli2021, Pauwels2022c} and not even generally in tasks where the receiver is interested only in a single property of the sender's input \cite{Vieira2022}. High-dimensional entanglement, decohered through irreversible encoding operations and subsequently measured in entangled projections of systems with different dimensions, can genuinely improve quantum communication.

However, demonstrating that using entanglement with local dimension greater than the dimension of the channel provides an advantage in establishing classical correlations is experimentally challenging. It requires both the implementation of several sophisticated quantum operations, in particular, ancillary qubits and controlled gates  \cite{Vieira2022, Pauwels2022c}, and overall fidelities high enough to outperform all conceivable quantum strategies based on qubit entanglement. The latter are typically high, posing a further challenge to experiments.

Here, we experimentally realize qubit messages assisted by four-dimensional entanglement that outperform all protocols based on qubit entanglement to assist the qubit channel. We achieve this in a scenario that combines two well-known tasks, namely quantum state discrimination \cite{barnett2008quantum, Bae2015} and quantum random access coding \cite{Ambainis2002, ambainis2009quantum}. In the experiment, contrasting alternative encodings of high-dimensional entanglement \cite{fickler2012quantum, martin2017quantifying, hu2020efficient,wang2018multidimensional, kues2017chip, ringbauer2022universal, fu2022experimental, goss2022high}, we expand the path mode of a single photon to encode four-qubit entanglement and use information interconversion between path and polarization to achieve the required compression operations and the necessary three-qubit entangled measurements. This approach allows us to circumvent well-known no-go theorems for entangled measurements and deterministic control gates on separate photons \cite{Lutkenhaus1999, Loock2004}. Our results demonstrate that high-dimensional entanglement, as compared to standard qubit entanglement, can enhance qubit messages in the single shot regime.

\section{Theory}
Consider a scenario where two parties, Alice and Bob, share an entangled state $\ket{\psi}_{AB}$ with local dimension $d$. The parties are connected by a qubit channel, allowing Alice to send messages to Bob. On each round, Alice privately selects a symbol $x$ and encodes it into a qubit message which is sent over the channel, possibly using her share of the entangled state in the process. 
Bob in turn selects a private symbol $y$, to which he associates a quantum measurement, used to decode Alice's message, whose outcome is denoted $b$. The experiment is described by a probability distribution, $p(b|x,y)$, called the correlations in the experiment. This is the simplest setting to study the enhancement of qubit communication by high-dimensional entanglement. 

The quantum encoding of Alice is represented by a completely positive trace-preserving (CPTP) map $\Lambda_x$ acting on her share of $\ket{\psi}_{AB}$. The joint state of the received message and the share of the entangled state held by Bob is therefore
\begin{equation} \label{eq:tau}
	\tau_x=\left(\Lambda_x\otimes \openone\right)[\psi_{AB}] \, .
\end{equation}
Bob's measurements are represented by a positive operator-valued measure $\{M_{b|y}\}$. The correlations in the experiment are given by Born's rule,
\begin{equation}\label{born}
	p(b|x,y)=\Tr\left(\tau_x M_{b|y}\right) \, .
\end{equation}

In this setting, one can consider various communication tasks and study the potential of entanglement to enhance the parties' performance on such tasks with qubit messages. A primary example is the task of state discrimination, where the parties wish to maximize the probability of Bob correctly identifying the symbol $x$ of Alice. There, entanglement can boost Bob's ability to guess Alice's ability to four symbols, instead of two without entanglement \cite{Bennett1992}. More generally, Bob has several different inputs $y$, which can be viewed as different \emph{questions} about the input. For example, in the task of random access coding, Alice's inputs are interpreted as a string of two bits, and Bob is asked to randomly guess one of the bits \cite{Ambainis2002}. The success rate of the protocol is then the average probability that Bob correctly guesses the bit. A broad range of tasks can be described as the parties being rewarded with $c_{bxy}$ points depending on Bob's answer $b$ to a specific question $y$ about the value of $x$. The success rate takes the form,
\begin{equation}
	p_{\rm succes}=\sum_{b,x,y}c_{bxy}p(b|x,y) \,.
\end{equation}

In general, the CPTP maps of Alice are associated with compression, from her part of the entangled state of dimension $d$ to a qubit. Compression implies a loss of information, i.e.~ decoherence of the global entanglement. The exception is when the dimension of the output of the channel is equal to the input, $d=2$; then Alice can use unitary encodings, $U_x$, which leave the joint states pure,
\begin{equation}\label{unitary}
	\ket{\tau_x}=U_x\otimes\openone \ket{\psi_{AB}}.
\end{equation} 
In all other cases, when the entanglement dimension is greater than that of the communication line, the final states $\tau_x$ are qubit-qu$d$it systems and entanglement is partially destroyed by the compression operations.

A priori, it may therefore appear natural to assume it is always optimal to choose entanglement of the same dimension as the channel, $d=2$. This intuition is indeed correct for the paradigmatic example of state discrimination, which is optimally achieved by dense coding \cite{Bennett1992}.

Nevertheless, in what follows, we show that this intuition is generally misleading. We describe a task \cite{Tavakoli2021} for which a protocol based on $d=4$ outperforms any possible quantum protocol based on qubit entanglement. Since such states cannot have a Schmidt number greater than two, the only potential advantage of using high-dimensional entanglement ($d>2$) lies in the fact the collection of states $\{\tau_x\}$ can span different subspaces of truly high-dimensional space. 

\subsection{Communication task}

We consider a task that combines state discrimination and random access coding. Alice has five symbols in her alphabet: $x=(x_1,x_2)\in\{0,1\}^2$ for random access coding and $x=F$ is an additional flag state for state discrimination. Bob has three inputs: $y=1,2$ for random access coding, and $y=P$ for state discrimination. Bob's output is $b=0,1$. We are given a state discrimination condition, namely that  when $y=P$, Bob can always tell $x=(x_1,x_2)$ apart from $x=F$, i.e. $p(1|x,y=P)=\delta_{F,x}$. This requires that $\tau_{x_1x_2}$ and $\tau_F$ are orthogonal states.

Given this promise, the goal is to maximize the average success probability of Bob retrieving Alice's $y$'th bit,
\begin{equation}
	P_{\rm RAC}=\frac{1}{8} \sum_{y=1,2} \sum_{b,x_1,x_2 =0,1} \delta_{b x_y} p(b|x_1x_2,y) , . \label{eq:PRAC}
\end{equation}

Owing to the state discrimination promise, without entanglement assistance, the parties cannot do better than random. Indeed, the qubit message has to be used entirely to ensure state discrimination, and their performance on the random access coding task is limited to $P_{\rm RAC}^{\rm Qubit\ ent.} = 0.5$. To what extent can entanglement improve their performance?

\subsection{Optimal strategies under exact state discrimination}

Here, we look at the optimal strategies for this task assuming qubit entanglement which is optimally manipulated via unitary encodings. We then give an explicit strategy based on four-dimensional entanglement and irreversible encodings that outperforms all such strategies. Here, we assume that the state-discrimination promise is exactly fulfilled. This is evidently an idealisation which is incompatible with experiment. In the next section, we show how it can be eliminated.

\subsubsection{Qubit-entanglement-based strategies} \label{sec:qubitstrategies}
Consider any protocol based on unitary encoding operations on a shared maximally entangled state, as in Eq.~\eqref{unitary}. The space of such states is contained in the set of four-dimensional quantum states. However, it is further constrained by the impossibility of reading out the value of $x$ by measuring Bob's particle. It is known that the state space corresponding to \eqref{unitary} is isomorphic to that of four-dimensional quantum systems when they are restricted to a real-valued Hilbert space, i.e.~vectors in $\mathbb{R}^4$ \cite{Pauwels2022c}. We can therefore work in this simpler picture.

Under the perfect state-discrimination condition $p(1|x,y=P)=\delta_{F,x}$, the four real four-dimensional states $\tau_{x_0x_1}$ in the success probability ~\eqref{eq:PRAC} are confined to a three-dimensional subspace, namely the space orthogonal of that of $\tau_F$. Thus, the success probability $P_{\rm RAC}^{}$ is equal to that of the Random Access Code with real three-dimensional communication, $P_{\rm RAC}^{\rm Qubit\ ent.}=P_{\rm RAC}^{\rm real~qutrit}$. The optimal (complex) qutrit strategy for this case is known to achieve $P_{\rm RAC}^{qutrit}= (5+\sqrt{5})/8 \approx 0.904$ \cite{Tavakoli2018}. Furthermore, because this optimal qutrit strategy can be represented already in a real-valued Hilbert space, it follows that 
\begin{equation}\label{limit}
	P_{\rm RAC}^{\rm Qubit\ ent.}=P_{\rm RAC}^{\rm real~qutrit}=P_{\rm RAC}^{\rm qutrit}=\frac{5+\sqrt{5}}{8} \approx 0.904.
\end{equation}

We may note that this strategy beats the naive one in which one first uses dense coding to send two bits over the channel and then classically decide the output. Such a dense coding strategy is limited by $P_{\rm RAC}^{\rm DC} = 7/8$, where three symbols are used for the RAC and one for fulfilling the state discrimination condition.

\subsubsection{Strategy with four-dimensional entanglement and irreversible encoding} \label{sec:4dstrategy}
We now show how to beat the limit \eqref{limit} using  two copies of an EPR state, $|\phi_2\rangle_{A_1B_1}\otimes |\phi_2\rangle_{A_2B_2}$ and irreversible encodings for Alice. Specifically, let Alice apply a unitary $U_x$ on her qubits, $A_1A_2$, and then discard $A_1$. The unitaries are
\begin{align}
	&  U_1 = \openone \otimes \openone, \quad U_2 = C_1 C_2 ,  \quad  U_3 = \openone \otimes X C_1 C_2  \nonumber \\
	& \qquad U_4 = \openone \otimes  Z  \qquad U_5 = \openone \otimes ZX C_2  \label{eq:U_RAC}
\end{align}
where $C_i$ is the CNOT gate with control on $A_i$. Now, the optimal measurements can be analytically determined \cite{Tavakoli2021}. Clearly, state discrimination requires that  $M_{1|3}= \rho_F$. For the remaining two settings, one can write the objective function as $P_{\rm RAC}=\sum_{x_0x_1y}c_{x_0x_1,y}\tr[\tau_{x_0x_1}M_y]+1/2$, where $c_{x_0x_1,y}=(\delta_{0x_y}-\delta_{1x_y})/16$ and $M_y = M_{1|y}-M_{2|y}$. We rearrange the sum as $P_{\rm RAC}=\sum_{y}\tr[ (\sum_{x_0x_1} c_{x_0x_1,y} \tau_{x_0x_1})M_y]+1/2$ and observe that $(\sum_{x_0x_1} c_{x_0x_1,y} \tau_{x_0x_1})$ are Hermitian for $y=1,2$. The optimal measurement operators $M_{1|y}$ ($M_{2|y}$) are projective and align with the positive (negative) eigenspaces of $ (\sum_{x_0x_1} c_{x_0x_1,y} \tau_{x_0x_1})$. 

The optimal measurements are entangled over $B_1B_2A_2$ and $y=1,2$ project onto two different three-dimensional subspaces, each spanned by three orthogonal qubit-ququart states, locally equivalent to $\ket{\phi_2}$. $y=P$ has projectors of rank two and six, each composed of orthogonal maximally entangled states. We give the explicit form of the optimal measurements for the encoding ~\eqref{eq:U_RAC} in Appendix~\ref{app:1}.

Given this protocol, the success probability is
\begin{equation}\label{limit2}
	P_{RAC}^{\rm Ent. ~qubit} = \frac{6+\sqrt{2}}{8} \approx 0.927,
\end{equation}
which exceeds the limitation $P_{\rm RAC}^{\rm Qubit\ ent.}$ in Eq.~\eqref{limit}.

\subsection{Removing perfect state discrimination} \label{sec:correcting}
To make the protocol realizable in experiment, the constraint of having a perfect state discrimination measurement, $y=P$, must be relaxed.  This means that we have to modify our benchmark, the bound $P_{\rm RAC}^{\rm Qubit\ ent.}$, which was based on the assumption of perfect state discrimination.

To this end, we again use the correspondence between qubit-entanglement-assisted qubit communication and real ququart communication. When state discrimination is imperfect, the states $\{\tau_{x_1x_2}\}$ will only approximately correspond to a three-dimensional subspace, because they need no longer be perfectly orthogonal to $\tau_F$. We now show how to bound the dimensional deviation of ${\tau_{x_1x_2}}$, from exactly spanning a three-dimensional space, by using the data obtained in the state discrimination measurement $y=P$. Once such a bound is obtained, the correction to the figure of merit can be computed using the framework of  ``almost qudit'' introduced in  Ref.~\cite{Pauwels2022}.  

We apply the third measurement ${M_{0|P},M_{1|P}}$ to the five states $\rho_x$ for $x=00,01,10,11,F$, now represented as real ququarts, yielding probabilities of the form  $
\tr\left(\rho_x M_{\delta_{x,F}|P}\right)=r_x$. Ideally, we want the discrimination probabilities to be noise-free, i.e.  $r_x= 1$. This would mean that the states $\rho_{00},\ldots,\rho_{11}$ are restricted to a three-dimensional real Hilbert space. However, we measure noisy data with only $r_x\approx 1$. 
We now aim to determine bounds on the dimensional deviation parameters $\epsilon_x$, which quantify how far the states deviate from some three-dimensional subspace $\Pi_3$. This corresponds to relations
\begin{equation}
	\tr\left(\rho_x \Pi_3\right)\geq 1-\epsilon_x.
\end{equation}
Naturally, $r_x=1$ implies $\epsilon_x=0$. When $r_x\approx 1$,  we define $\tilde{M}_{1|P}$ as the rank-1 projector onto the eigenvector of $M_{1|P}$ associated to its largest eigenvalue. Then, for $x=00,\ldots,11$,
\begin{align}\nonumber
	r_x&=\tr\left(\rho_x M_{0|P}\right) \\ \nonumber
	&=\tr\left(\rho_x\left(\openone-M_{1|P}\right)\right) \\\nonumber 
	&=\tr\left(\rho_x\left(\openone-M_{1|P}+\tilde{M}_{1|P}-\tilde{M}_{1|P}\right)\right)\\
	& = \tr\left(\rho_x \left(\openone-\tilde{M}_{1|P}\right)\right) +\tr\left(\rho_x \left(\tilde{M}_{1|P}-M_{1|P}\right)\right).
\end{align}
Thus,
\begin{equation}
	\tr\left(\rho_x \left(\openone-\tilde{M}_{1|P}\right)\right)=r_x-\tr\left(\rho_x \left(\tilde{M}_{1|P}-M_{1|P}\right)\right).
\end{equation}
Here $\openone-\tilde{M}_{1|P}$ is the three-dimensional projector $\Pi_3$. We can put an upper bound on the right-most-term and hence a lower bound on the RHS by noticing that only the subspace $\tilde{M}_{1|P}$ in the eigenspace of $M_{1|P}$ has a non-negative weight. Hence for every choice of states, it holds that
\begin{equation}
	\tr\left(\rho_x \left(\tilde{M}_{1|P}-M_{1|P}\right)\right) \leq 1-\norm{M_{1|P}}\leq 1-r_5,
\end{equation}
where the last inequality is saturated by aligning $\rho_F$ with  $\tilde{M}_{1|P}$. Thus, we can view the states $\rho_{00},\ldots,\rho_{11}$ as almost three-dimensional real-valued qutrits, with a dimensional deviation quantified by
\begin{equation}\label{step}
	\tr\left(\rho_x \left(\openone-\tilde{M}_{1|P}\right)\right)\geq r_x+r_5-1=1-\underbrace{(2-r_x-r_5)}_{=\epsilon_x}.
\end{equation}

With the parameters $\{\epsilon_x\}$ in hand, we can compute the correction to the bound \eqref{limit} using the numerical method from  Ref.~\cite{Pauwels2022}. This consists of finding the "almost-qutrit" states that maximize $P_{\rm RAC}$ for random measurements of dimension $D$. These are states in dimension $D=4$ that satisfy $\tr\left(\rho_x \Pi_3\right)\geq 1-\epsilon_x$, where $\Pi_3$ is a diagonal matrix with three ones and the rest zeros in the computational basis, to approximate the three-dimensional identity operator. This is a semidefinite program, since $P_{\rm RAC}$ is linear in the states for fixed measurements and $\tr\left(\rho_x \Pi_3\right)\geq 1-\epsilon_x$ is a linear constraint. We can solve it efficiently with standard solvers such as MOSEK \cite{mosek}. We also find the measurements that maximize $P_{\rm RAC}$ for these states, which is another semidefinite program. We iterate between these two steps until convergence. We repeated this process for $\sim 100$ random initial measurements and got the same $P_{\rm RAC}$ each time. This was based on the parameters $\{\epsilon_x\}$ estimated from the experiment (see below). The corrected benchmark on qubit-entanglement protocols becomes 
\begin{equation}\label{newlim}
	P_{\rm RAC}^{\rm Corr.}\leq 0.910.
\end{equation}


\section{Implementation}
\begin{figure*}[tbph]
	\begin{center}
		\includegraphics [width=1.9\columnwidth]{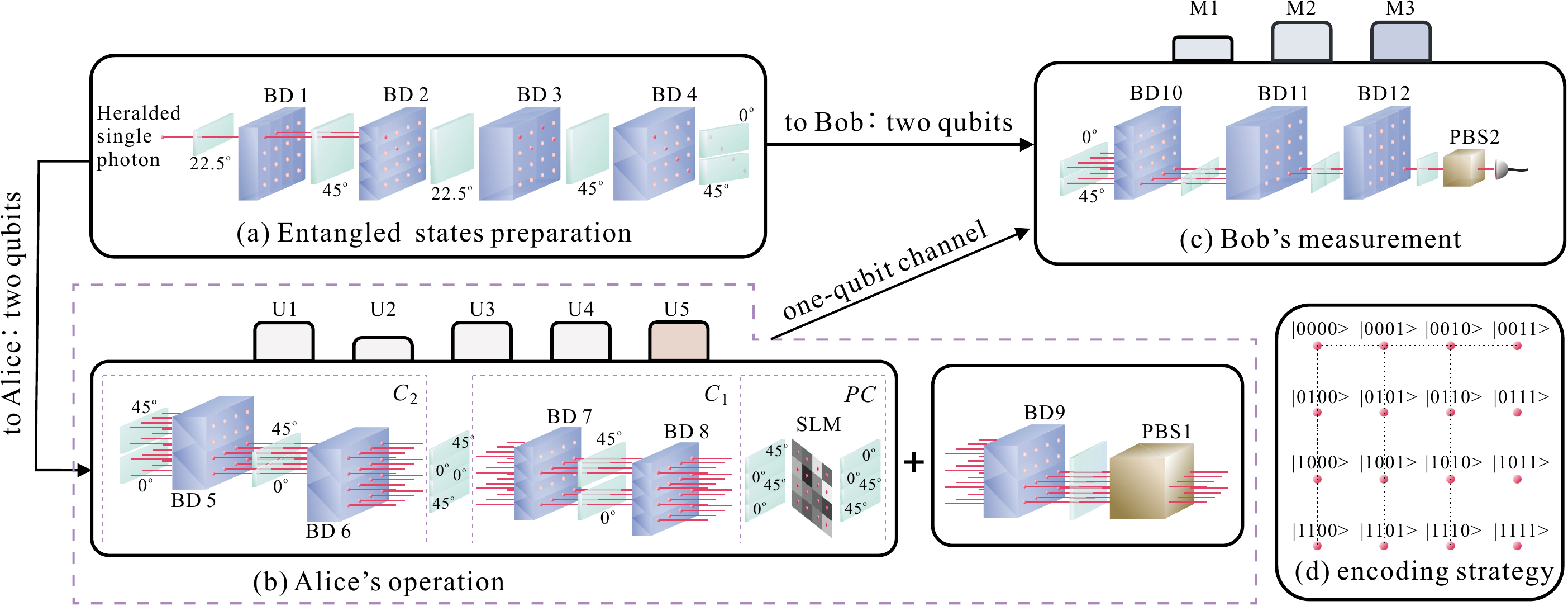}
	\end{center}
	\caption{\emph{Experimental setup.} (a) Preparation of two copies of the maximally entangled two-qubit state $\ket{\phi}_{A_1B_1}\otimes\ket{\phi}_{A_2B_2}$. (b) The compression operations on Alice's two qubits are realized via several building blocks which can act as Pauli gates, CNOT gates, and trace operations. (c)  HWPs, BDs, PBSs, and single photon detectors are used to jointly measure Bob's three qubits. (d) encoding of $A_1A_2B_1B_2$ in path modes. HWP: half-wave plate; BD: beam displacers; SLM: spatial light modulator; PBS: polarization beam splitter.}
	\label{fig:ExpSetup}
\end{figure*}

We implement the protocol, detailed in~\ref{sec:4dstrategy}, and show that we can exceed the success rate limitations of standard qubit entanglement. We use a four-qubit optical setup manipulating single-photon path-entanglement; see Figure~\ref{fig:ExpSetup}. It is composed of three main modules: (a) preparation and distribution of the entanglement, (b) realization of Alice's encoding operations and (c) Bob's decoding measurements. Below, we describe each module separately.

\subsection{State preparation} We expand the path modes of a single photon to encode high-dimensional entanglement, as shown in Figure~\ref{fig:ExpSetup} (a). The four qubits comprising the protocol state $\ket{\phi_2}_{A_1B_1}\otimes\ket{\phi_2}_{A_2B_2}$ are encoded into 16 path modes of a heralded single photon. A photon pair is generated using the process of spontaneous parametric down-conversion and is collected with single-mode fibers. With a pumping power of 1 mW and a 404 nm cw laser source, we collect $8\times 10^4$ pairs of 808 nm photons per second. When the idler photons trigger a single photon detector, the signal photons work as the target photons which are divided into 16 beams with beam displacers (BD1-4) and half-wave plates (HWPs). The BDs (16 mm $\times$ 16 mm $\times$ 8 mm, antireflection coating at 808 nm) are designed by stacking several polarizing beam splitters to introduce 4 mm (BD1, BD2) or 8 mm (BD3, BD4) separation between the horizontally and vertically polarized photons. To eliminate the diffraction effect, the diameter of each path mode is collimated to 1 mm. After the four BDs, the state of the 4-by-4 beam array takes the form $ \ket{\Phi}=\sum_k \alpha_k \exp{i\psi_k} \ket{k}_{A_1A_2B_1B_2}$ where $\alpha_k$ and $\psi_k$ are amplitudes and phases of the path modes, and $\ket{k}\in\{\ket{0000}, \ldots, \ket{1111}\}$. The states $\ket{0}$ and $\ket{1}$ for qubits $A_1$ and $A_2$ respectively, are encoded in the rows of the path modes, whereas for qubits $B_1$ and $B_2$ they are respectively encoded in the columns of the path modes (see Figure~\ref{fig:ExpSetup}(d)). Thus, operations acting on $A_1$ and $A_2$ can be implemented using BDs that direct photons in the vertical direction and thus have negligible effects on $B_1$ and $B_2$. The amplitudes $\{\alpha_k\}$ and relative phases $\{\psi_k\}$ need to be carefully controlled. For the former, the HWPs (from left to right in Figure~\ref{fig:ExpSetup}(a)) are successively set at $22.5\degree$, $45\degree$, $22.5\degree$, $45\degree$, $0\degree$, and $45\degree$. For the latter, we design a tailored structure using two HWPs and a spatial light modulator (SLM) with 16 separately controlled pixels to realize simultaneous control of the relative phases of 16 beam modes.

\subsection{Encoding operations} 
The information to be transmitted is encoded by performing the compression operations, which consist of a unitary gate in Eq.\eqref{eq:U_RAC} on Alice's two qubits $A_1A_2$ and a partial trace of $A_1$. Due to the flexibility of our single-photon encoding, the two-qubit controlled gates can be deterministically performed, which is not possible for the case on separate photons without postselection. Figure~\ref{fig:ExpSetup}(b) shows the assemblages of BDs and HWPs for realizing the CNOT gates $C_1$ and $C_2$. The core idea here is to properly recombine the path modes according to the status of the control qubit to transform the target qubit from path degree of freedom (DoF) to polarisation DoF and then performing Pauli $X$ or identity $I$ on the target qubit to achieve the CNOT gates. All the encoding operations can be implemented by cascading the CNOT assemblages, Pauli gates, and a partial trace operation (see Appendix~\ref{app:2}).

Our single-photon implementation has the advantage of high fidelity and flexibility to realize complicated multipartite processing. However, an accurate protocol demonstration requires Alice's devices to not influence Bob's two qubits. We carefully calibrated our setup by aligning the optical axes of BDs and HWPs to guide the photon in the vertical direction, thus eliminating the potential influence of Alice's compression operations on Bob's qubits. To estimate the residual effect, we calculated both the fidelity,  $\left(\Tr\sqrt{\sqrt{\rho_0}\rho\sqrt{\rho_0}}\right)^2$, and the quantum mutual information, $\Tr\left(\rho_0(\log\rho_0-\log\rho)\right)$, between the tomographically reconstructed states of Bob's share of the entangled state before, $\rho_0$, and after Alice's compression operations, $\rho$.  These are given in Table~\ref{table:F&QMI}. The average fidelity between before and after is $0.9991$, with a typical standard deviation of $0.0001$. The averaged quantum mutual information is $0.0018$, which confirms negligible correlations.

\begin{table}[h!]
	\centering
	\begin{tabular}{c|c|c|c|c|c}
		\hline
		& $U_1$ \quad & $U_2$ \quad & $U_3$ \quad & $U_4$ \quad & $U_5$ \quad \\
		\hline
		Fidelity & 0.9989\quad & 0.9991\quad & 0.9993\quad & 0.9991\quad & 0.9992\quad  \\
		\hline
		QMI & 0.0023\quad & 0.0019\qquad & 0.0014\quad & 0.0018\quad & 0.0016\quad \\
		\hline
	\end{tabular}
	\caption{The fidelity and quantum mutual information (QMI) of Bob's two-qubit states before and after Alice's compression operations.}
	\label{table:F&QMI}
\end{table}

\subsection{Measurements} The optimal decoding measurements, explicitly given in Appendix~\ref{app:1} are implemented in our setup as follows. Qubits $A_2B_1B_2$ are jointly projected onto well-chosen entangled subspaces encoded in the eight relevant photonic path modes. We use the configuration in Figure~\ref{fig:ExpSetup}(c) to implement all $8-$dimensional projections. These are implemented through separate projections on each of the pure entangled states comprising the degenerate measurement projectors. Thus, each lab measurement is effectively a projection onto a maximally entangled two-qubit space embedded in an eight-dimensional space. After the projective measurements are performed, the measurements $M_{1,2,P}$ can be achieved by postprocessing the experimental outcomes. See Appendix~\ref{app:2} for the angle settings for the HWPs in the measurement configurations corresponding to $M_{1,2,P}$.  We provide the overall experiment setup in Appendix~\ref{app:3} for completeness.

\section{Results} 

\begin{figure}
	\begin{center}
		\includegraphics [width=0.9\columnwidth]{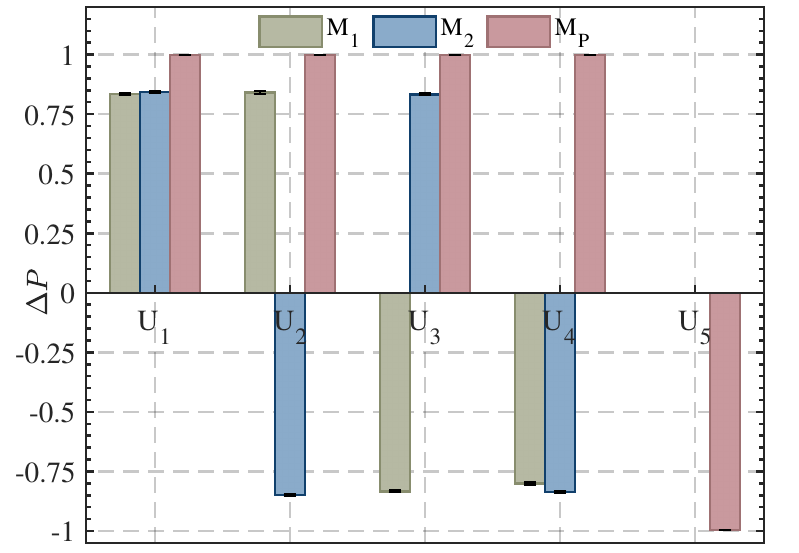}
	\end{center}
	\caption{\emph{Experimental estimate of the probability distribution.} Estimated expectation values $\Delta P_{xy}=p(b=0|x,y)-p(b=1|x,y)$ from measured coincidences for every input of Alice and Bob. The statistics of $M_{1,2}$ for the first four encoding operations $U_{1-4}$ of Alice contribute to the success probability of the random access code. The statistics of $M_P$ are used to determine the orthogonality of Alice's final encoding operation $U_5$ to the other encodings. See Appendix~\ref{app:5} for the experimental data.}
	\label{fig:M3Orthogonality}
\end{figure}

To estimate the correlations $p(b|x,y)$ in the experiment, we collected coincidences over a time of 8s for each pair of settings $(x,y)$, by resolving each rank-1 path-entangled projector for 1~s, and observed approximately $2\times 10^4$ events for each setting. By grouping the appropriate events, we obtain the estimate of $p(b|x,y)$, summarized in Figure~\ref{fig:M3Orthogonality}.

The data accurately agrees with the theoretical predictions. The estimated success rate in the task is 
\begin{equation}\label{measured}
	P_{\rm RAC}^{\rm exp}=0.9167\pm 0.0007,
\end{equation}
which is somewhat below the ideal theoretical value and well above the noise-free qubit-entanglement-based bound $P_{\rm RAC}^{\rm Qubit\ ent.}\approx 0.904$. A simple quality estimate of the experiments corresponds to assuming that with some probability the experiment runs ideally, reaching the theoretical limit in Eq.~\eqref{limit2}, and with the complement probability it outputs random results, based on sharing a maximally mixed state. Then, the measured value \eqref{measured} corresponds to a faithful implementation with over $97\%$ probability. This indicates a high overall accuracy in the implementation.
However, as is expected, we also observe small deviations from perfect state discrimination in the experimental data. These have to be taken into account to obtain a benchmark-to-beat for our experiment.

\subsection{Correction to the figure of merit and its violation}
\begin{figure}[t!]
	\begin{center}
		\includegraphics [width= 0.9\columnwidth]{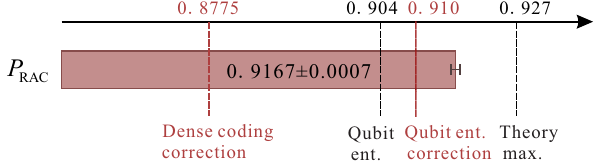}
	\end{center}
	\caption{\emph{Results.} The experimental result for $P_{RAC}$ (brown bar) and theoretical limits on qubit- and four-dimensional entanglement assistance without (black) and with (red) corrections implied by noisy state discrimination. The error bar represents one standard deviation.}
	\label{fig:2}
\end{figure}

We have measured the relative frequencies associated with the state discrimination measurement ($y=P$) applied to the five states. The successful outcomes in each case are found to correspond to 
\begin{align}
	r_1=0.9990\pm 0.0002\\
	r_2=0.9994\pm 0.0002\\
	r_3=0.9988 \pm 0.0002\\
	r_4=0.9993 \pm 0.0002\\
	r_5=0.9977\pm 0.0004.
\end{align}
Following section \ref{sec:correcting}, we use Eq.~\eqref{step} to compute the dimensional deviation parameters $\{\epsilon_x\}$, corresponding to how much our lab states deviate from a three-dimensional subspace. Since the relevant random variable for calculating $\epsilon_x$ is associated with the sum  $r_x+r_5$, its standard deviation is given by $\sqrt{\sigma_x^2+\sigma_5^2}$. All four relevant standard deviations become  $0.0004$ (they differ only from the fifth decimal). In order to safely account for the statistical fluctuations in $r_x+r_5$, we place each of the four random variables at five standard deviations below the mean value, i.e.~$\epsilon_x=2-\bar{r}_x-\bar{r}_5+5\sqrt{\sigma_x^2+\sigma_5^2}$.  The corresponding dimensional deviations are $\epsilon_1=0.0054$, $\epsilon_2=0.0049$, $\epsilon_3=0.0056$ and $\epsilon_4=0.0050$.  These parameters were used in section~\ref{sec:correcting} to arrive to Eq.~\eqref{newlim} which serves as the  noise-corrected theoretical upper bound on qubit-entanglement-based strategies becomes. This is still below the experimental value in Eq.~\eqref{measured}. The main results are illustrated in Figure~\ref{fig:2}.

\subsection{Statistical significance of the observed violation}

Finally, we compute the statistical significance of the experimental violation of the corrected bound $P_{\rm RAC}^{\rm Corr}$. To this end, we use the Hoeffding inequality \cite{hoeffding1994probability} to compute an upper bound on the probability of the violation being a consequence of the finite statistics, i.e.~a $p$-value estimate. We follow \cite{Gill2002}. Denote $x_i$, $y_i$ and $b_i$ the respective inputs of Alice and Bob and Bob's outcome in the $i$th experimental run. Consider the random variables
\begin{equation}
	(\hat{P}_{\rm RAC})_i = \sum_{xyb} c_{bxy}\frac{ \chi(b_i=b,x_i=x,y_i=y)}{p(x,y)},
\end{equation}
where $\chi(e)$ is the indicator function for the event $e$, i.e. $\chi(e)= 1$ if the event is observed and $\chi(e) = 0$ otherwise. The prior probabilities on the choice of settings in our experiment were chosen uniformly, $p(x,y) = 1/(2 \times 4) =1/8$.

We allow the random variable $(\hat{P}_{\rm RAC})_i$ to depend on past events, $j<i$, but not on future events, $j>i$. We define our estimator for the functional ${P}_{\rm RAC}$, from these random variables, $\hat{P}_{\rm RAC} = \frac{1}{N} \sum_{i=1}^N (\hat{P}_{\rm RAC})_i$, where $N \sim 4 \times 2 \times 20 \times 10^3$ is the total number of experimental rounds.

We then use the Azuma-Hoeffding inequality to express the $p$-value for our experiment. It implies a bound on the probability $p$ that our experimental results can be explained by qubit communication with qubit entanglement. Our figure of merit, $P_{\rm RAC}$ only takes into account the outcomes of  the RAC protocol, and the third measurement should be viewed as implying a constraint under which the success probability $P_{\rm RAC}$ is realized. In particular, the constraint implied by the third measurement in our experiment implies a constraint on the qubit-entanglement-based model equivalent to demanding that the observed value of $P_{\rm RAC}$ be explained by a real almost qutrit model.

We conclude that the probability that the observed value of $\hat{P}_{\rm RAC}$ can be explained by a qubit-entanglement-based model is bounded by \cite{Azuma1967}

\begin{align}
	p\Bigg(\frac{1}{N} \sum_{i=1}^N (\hat{P}_{\rm RAC})_i \geq &P_{\rm RAC}^{\rm Corr} + \mu \Bigg) \leq \exp \Bigg(-\frac{2N\mu^2}{(c+T)^2} \Bigg),
\end{align}
where $\mu=0.0067$ is the observed violation of the corrected qubit entanglement bound, $T=-0.09$  is the corrected qubit entanglement bound on $-P_{\rm RAC}$ (computed again by relaxing to almost qutrits, using the methods detailed in \ref{sec:correcting}) and $c\equiv \text{max}_{xy}c_{xy}/p(x,y)=1$ is the algebraic maximum. Owing to the significant violation magnitude, which is circa $40\%$ of the range between the lower limit  $P_{\rm RAC}^{\rm Corr}$, the upper limit $P_{\rm RAC}^{\rm Ent. qubit}$, and a large number of coincidences per setting, the probability of the null hypothesis is vanishingly small, $p\sim 10^{-8}$.

\section{Discussion}
By exploiting 16 path modes for a heralded single photon, we reported on a demonstration of the use of high-dimensional entanglement and irreversible encoding operations to boost the communication ability of qubits beyond any possible protocol based on unitary encodings that preserve entanglement. While the use of photonic path modes is particularly suitable for our proof-of-principle experiment, a natural next challenge is to achieve these advantages using separate physical carriers of quantum information. A central obstacle for this is the implementation of entangled measurements on separate photonic carriers, which is well-known to in general be impossible to achieve deterministically in linear optics without auxiliary photons. An interesting prospect, that perhaps may allow one to circumvent this issue, is the possibility of basing single-shot entanglement-assisted quantum communication protocols on the exclusive use of interference-free measurements  \cite{Piveteau2022, Piveteau2023, bakhshinezhad2024}.

Finally, we discuss an interesting approach, that is complementary to the one discussed in the main text, for the correlation advantages that we have studied here. This concerns the activation of a quantum advantage by high-dimensional entanglement and irreversible encodings. Specifically, there exist distributed tasks in which qubit-entanglement and qubit messages cannot outperform even two bits of classical communication, i.e.~the dense coding protocol. However, the use of high-dimensional entanglement activates this possibility \cite{Vieira2022}. In Appendix~\ref{app:6}, we identify two optimal (facet) tests of classical correlations featuring only a single measurement input. Interestingly, in such scenarios, the classical polytope coincides with the quantum set of correlations \cite{Frenkel2015}. Furthermore, given the isomorphism between qubit-entanglement-assisted qubit communication and real ququart communication \cite{Pauwels2022c}, the set of entanglement-assisted qubit correlations coincides with the polytope of two-bit correlations and hence entanglement cannot boost correlations beyond boosting the classical information capacity by dense coding \cite{Piveteau2022}. We prove that four-dimensional entanglement activates quantum correlations and that this can decrease the noise thresholds required for implementing these protocols. We provide an explicit gate decomposition in the form of a quantum circuit for implementing the activation phenomenon.

\medskip
\textbf{Acknowledgements} \par 
J.P is a FRIA grantee of the FRNS-F.R.S. (Belgium) and acknowledges funding from the QuantERA II Programme which has received funding from the European Union's Horizon 2020 research and innovation program under Grant Agreement No 101017733 and the F.R.S-FNRS Pint-Multi program under Grant Agreement R.8014.21, from the European Union’s Horizon Europe research and innovation program under the project "Quantum Security Networks Partnership" (QSNP, grant agreement No 101114043), from the F.R.S-FNRS through the PDR T.0171.22, from the FWO and F.R.S.-FNRS under the Excellence of Science (EOS) program project 40007526, from the FWO through the BeQuNet SBO project S008323N.
A.T. is supported by the Wenner-Gren Foundation and by the Knut and Alice Wallenberg Foundation through the Wallenberg Center for Quantum Technology (WACQT). E.Z.C. acknowledges funding by FCT/MCTES - Fundação para a Ciência e a Tecnologia (Portugal) - through national funds and when applicable co-funding by EU funds under the project UIDB/50008/2020. The USTC group was supported by the NSFC (No.~12374338, No.~12174367, No.~12204458, and No.~12350006), the Innovation Program for Quantum Science and Technology (No. 2021ZD0301200), the Fundamental Research Funds for the Central Universities, Anhui Provincial Natural Science Foundation (No. 2408085JX002),  Anhui Province Science and Technology Innovation Project (No. 202423r06050004), China Postdoctoral Science Foundation (BX2021289 and 2021M700138). This work was partially carried out at the USTC Center for Micro and Nanoscale Research and Fabrication.

Funded by the European Union. Views and opinions expressed are however those of the author(s) only and do not necessarily reflect those of the European Union, which cannot be held responsible for them.

Y. G. and H. T. contributed equally to this work.

\medskip
\textbf{Data Availability Statement} \par 
The data that support the findings of this study are available from the corresponding author upon reasonable request.

\bibliography{Reference_beyondDC}

\appendix

\onecolumngrid

\section{Optimal decoding measurements four-dimensional entanglement RAC strategy} \label{app:1} \par
Here, we give the explicit form of the optimal decoding measurements for the four-dimensional entanglement RAC strategy, found by solving the eigenvalue problem detailed in Subsection 2.2.2 of the main text. The optimal decoding measurements are given by

\begin{align}\label{eq:M1}
	M_{1} = \left[ \begin{array}{cccccccc}
		0.3536 & 0 & -0.1464 & 0 & 0 & 0.8536 & 0 & 0.3536 \\
		0 & 0.5 & 0 & 0 & 0 & 0 & 0.5 & 0 \\
		-0.1464 & 0 & -0.3536 & 0 & 0 & -0.3536 & 0 & 0.8536 \\
		0 & 0 & 0 & -0.5 & -0.5 & 0 & 0 & 0 \\
		0 & 0 & 0 & -0.5 & -0.5 & 0 & 0 & 0 \\
		0.8536 & 0 & -0.3536 & 0 & 0 & -0.3536 & 0 & -0.1464 \\
		0 & 0.5 & 0 & 0 & 0 & 0 & 0.5 & 0 \\
		0.3536 & 0 & 0.8536 & 0 & 0 & -0.1464 & 0 & 0.3536 \end{array} \right],
\end{align}

\begin{align}\label{eq:M2}
	M_{2} = \left[ \begin{array}{cccccccc}
		-0.3536 & 0 & -0.1464 & 0 & 0 & 0.8536 & 0 & -0.3536 \\
		0 & -0.5 & 0 & 0 & 0 & 0 & -0.5 & 0 \\
		-0.1464 & 0 & 0.3536 & 0 & 0 & 0.3536 & 0 & 0.8536 \\
		0 & 0 & 0 & 0.5 & 0.5 & 0 & 0 & 0 \\
		0 & 0 & 0 & 0.5 & 0.5 & 0 & 0 & 0 \\
		0.8536 & 0 & 0.3536 & 0 & 0 & 0.3536 & 0 & -0.1464 \\
		0 & -0.5 & 0 & 0 & 0 & 0 & -0.5 & 0 \\
		-0.3536 & 0 & 0.8536 & 0 & 0 & -0.1464 & 0 & -0.3536 \end{array} \right].
\end{align}

We can choose the eigenstates of $M_1$  as
\begin{align} \label{ensemble1}
	&(0.2895, 0, -0.5804, -0.2332, -0.2332, -0.6497, 0, 0.2201), \notag \\
	&(0.4636, 0, 0.3675, 0.3395, 0.3395, -0.5035, 0, -0.4073), \notag \\
	&(0.1564, 0, 0.4526, -0.5747, -0.5747, -0.0338, 0, -0.33), \notag  \\
	&(0, 0, 0, 1, -1, 0, 0, 0)/\sqrt{2}, \notag  \\
	&(0, 1, 0, 0, 0, 0, -1, 0)/\sqrt{2}, \notag \\
	&(-0.4730, 0.5312, 0.2332, 0, 0, -0.3737, 0.5312, 0.1339), \notag \\
	&(0.5440, 0.4667, -0.2591, 0, 0, 0.4260, 0.4667, -0.1411), \notag \\
	&(0.3964, -0.0066, 0.4491, 0, 0, 0.0462, -0.0066, 0.7993),
\end{align}
with respective eigenvalues $(-1, -1, -1, 0, 0, 1, 1, 1)$. The eigenstates of $M_2$ can be chosen as
\begin{align} \label{ensemble2}
	&(0.8032, 0.0337, 0.2070, 0, 0, -0.5562, 0.0227, 0.0399), \notag \\
	&(0.1760, 0.0022, -0.5295, 0, 0, 0.1162, 0.0022, 0.8217), \notag \\
	&(0.0263, -0.7067, 0.0050, 0, 0, -0.0175, -0.7067, 0.0039), \notag  \\
	&(0, 0, 0, 1, -1, 0, 0, 0)/\sqrt{2}, \notag  \\
	&(0, 1, 0, 0, 0, 0, -1, 0)/\sqrt{2}, \notag \\
	&(0.0325, 0, 0.0049, 0.7059, 0.7059, 0.0480, 0, -0.0106), \notag \\
	&(0.4306, 0, -0.6273, -0.0237, -0.0237, 0.3492, 0, -0.5459), \notag \\
	&(-0.3697, 0, -0.5322, 0.0345, 0.0345, -0.7433, 0, -0.1586),
\end{align}
with respective eigenvalues $(-1, -1, -1, 0, 0, 1, 1, 1)$. For the third measurement $M_P$, we can choose the following projectors
\begin{align} \label{ensemble3}
	&(0, 0, 0, 1, 1, 0, 0, 0)/\sqrt{2}, \notag \\
	&(0, 1, 0, 0, 0, 0, -1, 0)/\sqrt{2}, \notag \\
	&(1, 0, 0, 0, 0, 1, 0, 0)/\sqrt{2}, \notag  \\
	&(0, 0, 0, 1, -1, 0, 0, 0)/\sqrt{2}, \notag  \\
	&(0, 1, 0, 0, 0, 0, 1, 0)/\sqrt{2}, \notag \\
	&(1, 0, 0, 0, 0, -1, 0, 0)/\sqrt{2}, \notag \\
	&(0, 0, 1, 0, 0, 0, 0, 1)/\sqrt{2}, \notag \\
	&(0, 0, 1, 0, 0, 0, 0, -1)/\sqrt{2},
\end{align}

\begin{figure*}[tbph]
	\begin{center}
		\includegraphics [width=0.5\columnwidth]{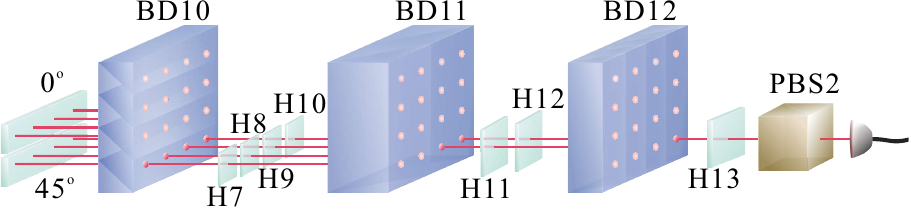}
	\end{center}
	\caption{\emph{Bob's decoding measurement apparatus.} All projectors $M_{yi}$ for $y=1, 2, P$ and $i=1,\ldots,8$ can be achieved by properly setting the angles of the HWPs (H7-H13), which are shown in Tabs~(\ref{table:SM2}-\ref{table:SM4}).}
	\label{fig:smmeasurement}
\end{figure*}


We use the configuration in Fig.~\ref{fig:smmeasurement} (also see Fig.~2(c) of the main text) to realize the projective measurements onto the eigenstates in Eq~(\ref{ensemble1}-\ref{ensemble3}). We denote each rank-1 projector as $M_{yi}$ for $i=1,\ldots,8$. Each projection is achieved by configuring the HWPs H7-H13 from left to right. The angle settings for these HWPs are listed in Tabs~(\ref{table:SM2}-\ref{table:SM4}). Although the scheme is deterministic, our method to implement the measurements $M_{y}$ for $y=1, 2, P$ makes our experiment suffer from the postselection problem \cite{Tavakoli2024}. Deterministic implementation of a general 8-dimensional measurement can be achieved by constructing a quantum random walk-based setup~\cite{Kurzy13prl,YG21prl} to transform the measurement into the one consisting of computational bases. However, this may severely decrease the overall fidelity and can hardly be adopted in our experiment using state-of-the-art techniques.

\begin{table*}[!htbp]
	\centering
	\caption{The angle settings of HWPs to realize Bob's decoding measurement $M_1$.}
	\begin{tabular}{c|c|c|c|c|c|c|c}
		\hline
		$M_1$ &\quad $H7$ \quad &\quad $H8$ \quad &\quad $H9$ \quad &\quad $H10$ \quad &\quad $H11$ \quad &\quad $H12$ \quad & \quad $H13$ \quad\\
		\hline
		\quad $M_{11}$ \quad & $96.63\degree$ & $90\degree$ & $0\degree$ & $160.69\degree$ & $61.94\degree$ & $18.54\degree$ & $24.48\degree$ \\
		\hline
		\quad $M_{12}$ \quad & $85.57\degree$ & $0\degree$ & $90\degree$ & $10.29\degree$ & $71.71\degree$ & $13.22\degree$ & $23.29\degree$ \\
		\hline
		\quad $M_{13}$ \quad & $133.86\degree$ & $90\degree$ & $90\degree$ & $43.47\degree$ & $75.3\degree$ & $28.86\degree$ & $19.72\degree$ \\
		\hline
		\quad $M_{14}$ \quad & $0\degree$ & $135\degree$ & $45\degree$ & $0\degree$ & $90\degree$ & $0\degree$ & $22.5\degree$ \\
		\hline
		\quad $M_{15}$ \quad & $90\degree$ & $135\degree$ & $45\degree$ & $90\degree$ & $45\degree$ & $45\degree$ & $22.5\degree$ \\
		\hline
		\quad $M_{16}$ \quad & $45\degree$ & $111.18\degree$ & $116.92\degree$ & $135\degree$ & $68.78\degree$ & $36.17\degree$ & $24.84\degree$ \\
		\hline
		\quad $M_{17}$ \quad & $45\degree$ & $60.07\degree$ & $55.16\degree$ & $135\degree$ & $62.18\degree$ & $39.58\degree$ & $23.83\degree$ \\
		\hline
		\quad $M_{18}$ \quad & $45\degree$ & $94.39\degree$ & $179.24\degree$ & $45\degree$ & $67.48\degree$ & $2.7\degree$ & $18.8\degree$ \\
		\hline
	\end{tabular}
	\label{table:SM2}
\end{table*}

\begin{table*}[!htbp]
	\centering
	\caption{The angle settings of HWPs to realize Bob's decoding measurement $M_2$.}
	\begin{tabular}{c|c|c|c|c|c|c|c}
		\hline
		$M_2$ &\quad $H7$ \quad &\quad $H8$ \quad &\quad $H9$ \quad &\quad $H10$ \quad &\quad $H11$ \quad &\quad $H12$ \quad & \quad $H13$ \quad\\
		\hline
		\quad $M_{21}$ \quad & $45\degree$ & $97.77\degree$ & $93.19\degree$ & $45\degree$ & $45\degree$ & $8.18\degree$ & $17.08\degree$ \\
		\hline
		\quad $M_{22}$ \quad & $0\degree$ & $135\degree$ & $45\degree$ & $0\degree$ & $45\degree$ & $45\degree$ & $22.5\degree$ \\
		\hline
		\quad $M_{23}$ \quad & $0\degree$ & $45\degree$ & $135\degree$ & $0\degree$ & $90\degree$ & $0\degree$ & $22.5\degree$ \\
		\hline
		\quad $M_{24}$ \quad & $90.88\degree$ & $0\degree$ & $0\degree$ & $0.43\degree$ & $89.89\degree$ & $1.19\degree$ & $22.49\degree$ \\
		\hline
		\quad $M_{25}$ \quad & $136.31\degree$ & $0\degree$ & $0\degree$ & $46.31\degree$ & $67.61\degree$ & $22.58\degree$ & $22.46\degree$ \\
		\hline
		\quad $M_{26}$ \quad & $46.53\degree$ & $90\degree$ & $0\degree$ & $43.5\degree$ & $56.15\degree$ & $33.62\degree$ & $22.5\degree$ \\
		\hline
		\quad $M_{27}$ \quad & $45\degree$ & $178.68\degree$ & $-6.86\degree$ & $45\degree$ & $-3.51\degree$ & $34.46\degree$ & $27.96\degree$ \\
		\hline
		\quad $M_{28}$ \quad & $45\degree$ & $45.09\degree$ & $46.88\degree$ & $45\degree$ & $46.18\degree$ & $41.86\degree$ & $22.46\degree$ \\
		\hline
	\end{tabular}
	\label{table:SM3}
\end{table*}

\begin{table*}[!htbp]
	\centering
	\caption{The angle settings of HWPs to realize Bob's decoding measurement $M_P$.}
	\begin{tabular}{c|c|c|c|c|c|c|c}
		\hline
		$M_P$ &\quad $H7$ \quad &\quad $H8$ \quad &\quad $H9$ \quad &\quad $H10$ \quad &\quad $H11$ \quad &\quad $H12$ \quad & \quad $H13$ \quad\\
		\hline
		\quad $M_{P1}$ \quad & $90\degree$ & $0\degree$ & $0\degree$ & $0\degree$ & $90\degree$ & $0\degree$ & $22.5\degree$ \\
		\hline
		\quad $M_{P2}$ \quad & $0\degree$ & $0\degree$ & $0\degree$ & $0\degree$ & $90\degree$ & $0\degree$ & $22.5\degree$ \\
		\hline
		\quad $M_{P3}$ \quad & $0\degree$ & $45\degree$ & $45\degree$ & $0\degree$ & $45\degree$ & $45\degree$ & $22.5\degree$ \\
		\hline
		\quad $M_{P4}$ \quad & $0\degree$ & $45\degree$ & $135\degree$ & $0\degree$ & $45\degree$ & $45\degree$ & $22.5\degree$ \\
		\hline
		\quad $M_{P5}$ \quad & $45\degree$ & $90\degree$ & $0\degree$ & $0\degree$ & $90\degree$ & $45\degree$ & $22.5\degree$ \\
		\hline
		\quad $M_{P6}$ \quad & $45\degree$ & $0\degree$ & $0\degree$ & $0\degree$ & $90\degree$ & $45\degree$ & $22.5\degree$ \\
		\hline
		\quad $M_{P7}$ \quad & $0\degree$ & $0\degree$ & $0\degree$ & $45\degree$ & $45\degree$ & $0\degree$ & $45\degree$ \\
		\hline
		\quad $M_{P8}$ \quad & $0\degree$ & $0\degree$ & $0\degree$ & $135\degree$ & $45\degree$ & $0\degree$ & $22.5\degree$ \\
		\hline
	\end{tabular}
	\label{table:SM4}
\end{table*}


\section{Details on Alice's encoding operations} \label{app:2} \par
\begin{figure}[t!]
	\begin{center}
		\includegraphics [width=0.5\columnwidth]{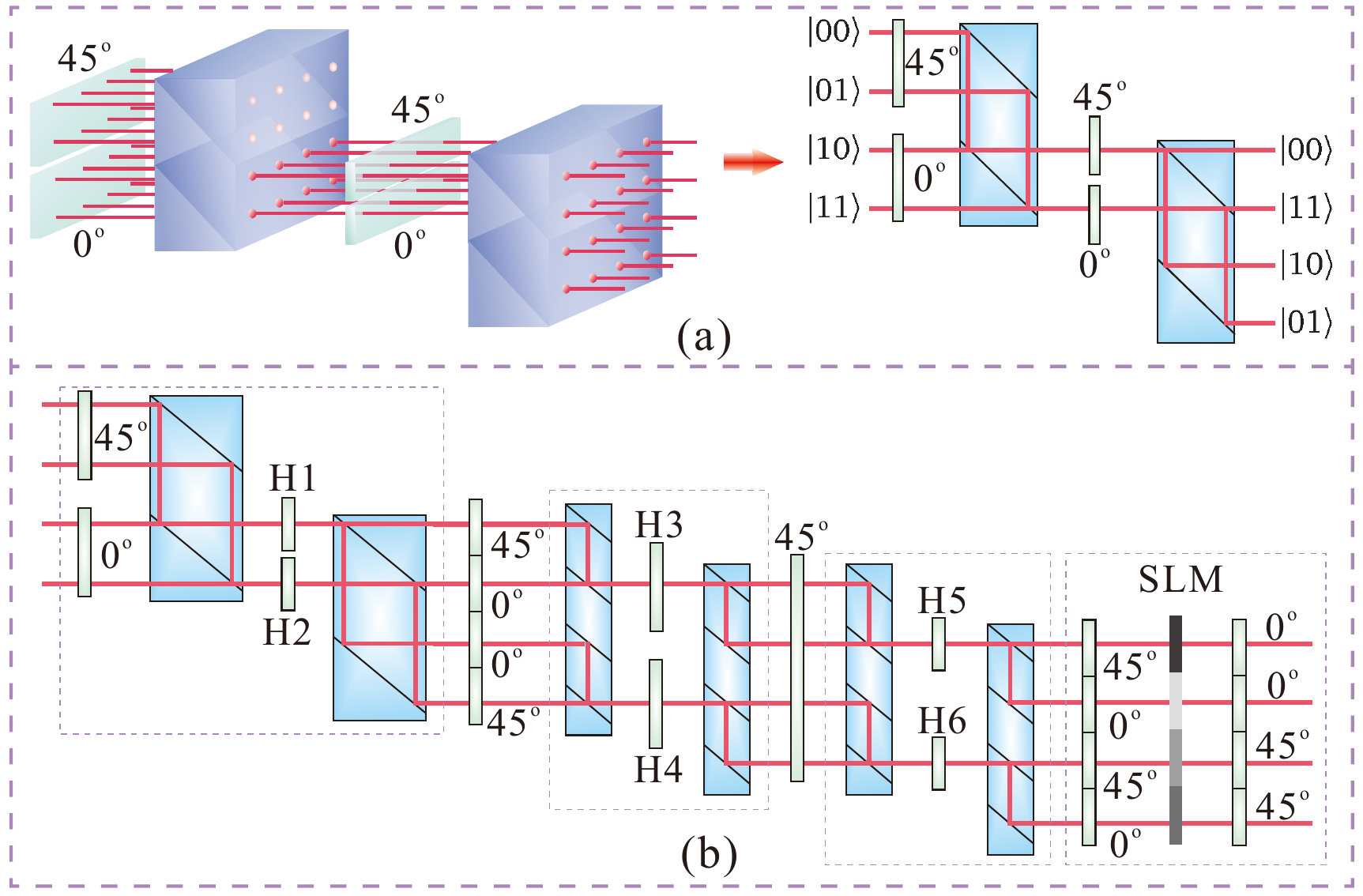}
	\end{center}
	\caption{\emph{Alice's encoding gates.} (a) The sub-module used to realize the gate $CNOT_2$ and its action on $A_1A_2$. (b) Setup for realizing all the unitary gates in Eq.(7) of the main text. CNOT gates and Pauli gates are achieved with four building blocks, via control of HWPs (H1-H6) and the SLM.}
	\label{fig:AliceEncoding}
\end{figure}

In this section, we provide details about the encoding operations, which consist of a unitary gate and a partial trace. The core here is to realize the CNOT gate in Eq.~(7) of the main text. The sub-module to realize the gate $C_{2}$  can be treated as Fig.~\ref{fig:AliceEncoding}(a) when operating on $A_1A_2$. A column of four beam modes in Fig.~\ref{fig:AliceEncoding}(a) is used to encode the computational basis of $A_1A_2$. After the assemblage that consists of two BDs and four HWPs, the modes encoding $\ket{00}_{A_1A_2}$ and $\ket{10}_{A_1A_2}$ remain unchanged, while the modes encoding $\ket{01}_{A_1A_2}$ and $\ket{11}_{A_1A_2}$ are swapped. This achieves the $C_{2}$ gate. Using this method, we design the setup in Fig.~\ref{fig:AliceEncoding}(b) for realizing the unitary gates in Eq.~(7) of the main text. It consists of four building blocks, each with two alternative settings achieved by setting the angles of the HWPs or applying voltages on the SLM. The settings are $C_{2}$ and $\openone\otimes\openone$ for the first block, $C_{1}$ and $\openone\otimes\openone$ for the second block,  $\openone\otimes\sigma_x$ and $\openone\otimes\openone$ for the third block, and  $\openone\otimes\sigma_z$ and $\openone\otimes\openone$ for the fourth block. The angle settings of the HWPs used to realize Alice's encoding unitary gates are listed in Tab.~\ref{table:UnitaryHWP}. Loss is a critical challenge in quantum communication tasks. In practical scenarios, the primary source of loss originates from lossy quantum channels. The encoding operation in our scheme significantly enhances loss tolerance by reducing the number of particles required to traverse a lossy channel. Compared to the standard approach utilizing qubit entanglement and qubit channels, our experiment increases the success probability of the random access code from 0.904 to approximately 0.917. This improvement demonstrates that our experiment can tolerate up to $1.4\%$ photon loss before its advantage is negated. Notably, the qubit entanglement approach also faces loss issues and both schemes depend on the transmission of single qubits.

\begin{table}[h!]
	\centering
	\caption{The angle settings of HWP1-6 to realize Alice's encoding unitary gates.}
	\begin{threeparttable}
		\begin{tabular}{c|c|c|c|c|c|c}
			\hline
			&\quad $H1$ \quad&\quad $H2$ \quad &\quad $H3$ \quad &\quad $H4$ \quad &\quad $H5$ \quad &\quad $H6$ \quad\\
			\hline
			\quad $U_1$ \quad & $45\degree$ & $45\degree$ & $45\degree$ & $45\degree$ & $45\degree$ & $45\degree$ \\
			\hline
			\quad $U_2$ \quad & $45\degree$ & $0\degree$ & $45\degree$ & $0\degree$ & $45\degree$ & $45\degree$ \\
			\hline
			\quad $U_3$ \quad & $45\degree$ & $0\degree$ & $45\degree$ & $0\degree$ & $0\degree$ & $0\degree$ \\
			\hline
			\quad $U_4$ \quad & $45\degree$ & $45\degree$ & $45\degree$ & $45\degree$ & $45\degree$ & $45\degree$ \\
			\hline
			\quad $U_5$ \quad & $45\degree$ & $0\degree$ & $45\degree$ & $45\degree$ & $0\degree$ & $0\degree$ \\
			\hline
		\end{tabular}
		\begin{tablenotes}
			\item[*] For $U_4=\openone \otimes Z$, the HWP angle settings in the table can only realize $\openone \otimes \openone$. The required $Z$ gate is achieved by properly setting the SLM.
		\end{tablenotes}
	\end{threeparttable}
	\label{table:UnitaryHWP}
\end{table}

\begin{figure}[t!]
	\begin{center}
		\includegraphics [width=0.9\columnwidth]{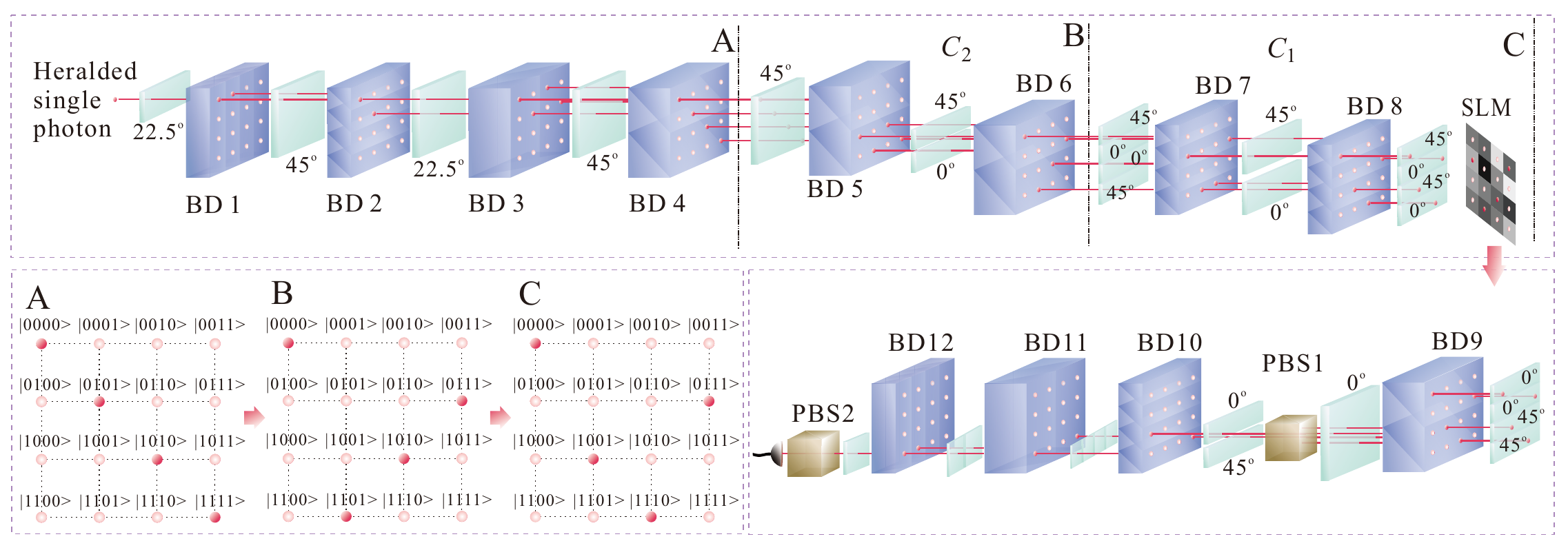}
	\end{center}
	\caption{\emph{Overall experimental setup.} The angle settings of the HWPs in the figure realize the gates $C_{2}$ and $C_{1}$. The bottom left panel shows the distributions of photon path modes in planes A, B, and C.}
	\label{fig:setup}
\end{figure}

\section{Overall experimental setup} \label{app:3}

In Fig.~\ref{fig:setup}, we provide the overall experimental setup for completeness.

\section{Measured correlations} \label{app:5} 
By grouping the appropriate outcomes for given $(x,y)$, we obtain an estimator for the probabilities, $p(b|x,y)$. The corresponding correlations are listed in Tab.~\ref{table:Mea_corr}. The error bars, calculated from 100 simulations of Poisson statistics, are smaller than the corresponding correlation with at least one order of magnitude. These correlations are grouped to estimate the success probability of the random access code game and to determine the orthogonality of $U_5$ to $U_{1-4}$; see Fig. 3 of the main text. 

\begin{table*}[!htbp]
	\centering
	\caption{Measured correlations for each compression operation and measurement setting $p(b|x,y)$.}
	\begin{tabular}{c|c|c|c|c|c|c|c|c}
		\hline
		\multicolumn{9}{c}{Measured correlations for $M_1$}   \\
		\hline
		& \quad $M_{11}$ \quad & \quad $M_{12}$ \quad & \quad $M_{13}$ \qquad & \quad $M_{14}$ \quad & \quad $M_{15}$ \qquad & \quad $M_{16}$ \quad &\quad $M_{17}$ \qquad & \quad $M_{18}$ \quad \\
		\hline
		$U_1$ & 0.0498 & 0.0194 & 0.0137 & 0 & 0.0001 & 0.3209 & 0.1696 & 0.4264 \\
		\hline
		$U_2$ & 0.0462 & 0.0229 & 0.0101 & 0.0001 & 0.0006 & 0.3000 & 0.3493 & 0.3608 \\
		\hline
		$U_3$ & 0.4303 & 0.4460 & 0.0394 & 0.0006 & 0.0013 & 0.0115 & 0.0094 & 0.0615 \\
		\hline
		$U_4$ & 0.2816 & 0.1426 & 0.4764 & 0.0002 & 0 & 0.0064 & 0.0060 & 0.0868 \\
		\hline
		$U_5$ & 0.0039 & 0.0011 & 0.0001 & 0.4997 & 0.4915 & 0.0009 & 0.0025 & 0.0001 \\
		\hline
		\multicolumn{9}{c}{Measured correlations for $M_2$}   \\
		\hline
		\quad & \quad $M_{21}$ \quad & \quad $M_{22}$ \quad & \quad $M_{23}$ \qquad & \quad $M_{24}$ \quad & \quad $M_{25}$ \quad & \quad $M_{26}$ \quad &\quad $M_{27}$ \quad & \quad $M_{28}$ \quad \\
		\hline
		$U_1$ & 0.0453 & 0.0039 & 0.0294 & 0.0001 & 0 & 0.0002 & 0.4958 & 0.4253 \\
		\hline
		$U_2$ & 0.2582 & 0.4899 & 0.1758 & 0.0006 & 0.0001 & 0.0001 & 0.0004 & 0.0747 \\
		\hline
		$U_3$ & 0.0485 & 0.0005 & 0.0335 & 0.0010 & 0.0005 & 0.5048 & 0.0006 & 0.4105 \\
		\hline
		$U_4$ & 0.4499 & 0.0023 & 0.4658 & 0 & 0.0001 & 0.0002 & 0.0022 & 0.0794 \\
		\hline
		$U_5$ & 0.0002 & 0.0005 & 0 & 0.4865 & 0.5120 & 0.0004 & 0 & 0.0001 \\
		\hline
		\multicolumn{9}{c}{Measured correlations for $M_P$}   \\
		\hline
		\quad & \quad $M_{P1}$ \quad & \quad $M_{P2}$ \quad & \quad $M_{P3}$ \qquad & \quad $M_{P4}$ \quad & \quad $M_{P5}$ \quad & \quad $M_{P6}$ \quad &\quad $M_{P7}$ \quad & \quad $M_{P8}$ \quad \\
		\hline
		$U_1$ & 0.0006 & 0.5025 & 0.0004 & 0.0003 & 0.0006 & 0.0012 & 0.4938 & 0.0005 \\
		\hline
		$U_2$ & 0.0001 & 0.1252 & 0.5096 & 0.0005 & 0.0001 & 0.1247 & 0.1203 & 0.1194 \\
		\hline
		$U_3$ & 0.5039 & 0.1188 & 0.0002 & 0.0002 & 0.0009 & 0.1212 & 0.1268 & 0.1280 \\
		\hline
		$U_4$ & 0.0001 & 0.0011 & 0.0006 & 0.0006 & 0.0001 & 0.5114 & 0.0010 & 0.4850 \\
		\hline
		$U_5$ & 0.0007 & 0.0001 & 0.0011 & 0.5110 & 0.4867 & 0.0001 & 0 & 0.0002 \\
		\hline
	\end{tabular}
	\label{table:Mea_corr}
\end{table*}

\section{Circuit proposal for activation of a quantum-over-classical advantage} \label{app:6} 
In a fully classical communication scenario, in which shared entanglement is replaced by a shared classical random variable and the noise-free quantum channel is replaced with a noise-free classical channel, it is well-known that the set of classical correlations forms a polytope \cite{Gallego2010}. The facets of this polytope can therefore be viewed as optimal tests of classicality, in the sense that they are the strictest possible criteria that all classical communication models must obey. Interestingly, it was shown in Ref.~\cite{Frenkel2015} that when Bob performs a fixed measurement (i.e.~$y$ has only one value) then the set of quantum correlations without entanglement assistance is equivalent to the classical set, namely said polytope.  That is, quantum $d$-dimensional communication does not improve on classical $d$-dimensional communication.  In the context of entanglement assistance, it is known from \cite{Pauwels2022c} that the set of effective states given to Bob over a qubit channel assisted by a two-qubit maximally entangled state is isomorphic to the set of real-valued four-dimensional single quantum systems. Naturally, this is a subset of the set of all four-dimensional quantum systems. Consequently, putting the results together, qubit messages and qubit entanglement cannot beat the classical models with two bits of communication in any task in which Bob has a single input. 

However, if the parties share higher-dimensional entanglement, the isomorphism with real-valued four-dimensional systems no longer holds. Thus, it may be possible to violate classical limitations by exploiting such entanglement. We have investigated this by first identifying facets of the classical polytope with two bits of communication. Specifically, we solve the facet enumeration problem with a single input on Bob using the polytope slicing method in \cite{Pal2009,Jesus2023}. By $(X,d,Y,B)$ we denote a general prepare-and-measure scenario with $X$ preparations, messages of dimension $d$, $Y$ measurement choices and $B$ outcomes for these measurements. We find a total of five classes of facets in (6,4,1,6). For two of those classes, we did not find any quantum advantage up to five-dimensional entanglement. Two of them are part of two families of facet inequalities valid for $(d+1,d,1,d+2)$ scenarios, where $d \geq 2$, proposed originally in \cite{Vieira2022}. We call the latter $I_{d+1,d,d+2}^{(1)}$ and $I_{d+1,d,d+2}^{(2)}$. We could check up to $d=4$ that no new classes of facets emerge. 

The most interesting findings are the three new facets in the $(6,4,1,6)$ scenario. These are inequalities of the form
\begin{equation}
	\mathcal{F} = \sum_{b=1}^6\sum_{x=1}^6 c_{b,x} p(b|x) \leq C \, ,
\end{equation}
\begin{align}
	&c^{(1)} = \begin{bmatrix}
		-1 & 0 & 0 & 0 & 0 & 1 \\
		-1 & 0 & 0 & 0 & 1 & 0 \\
		-1 & 0 & 0 & 1 & 0 & 0 \\
		-1 & 0 & 1 & 0 & 0 & 0 \\
		-1 & 1 & 0 & 0 & 0 & 0 \\
		0 & 0 & 0 & 0 & 0 & 0
	\end{bmatrix} \, ,
	& c^{(2)} = \begin{bmatrix}
		-1 & -1 & 0 & 0 & 0 & 1 \\
		-1 & -1 & 0 & 0 & 1 & 0 \\
		-1 & -1 & 0 & 1 & 0 & 0 \\
		-1 & 0 & 1 & 0 & 0 & 0 \\
		0 & -1 & 1 & 0 & 0 & 0 \\
		0 &  0 & 0 & 0 & 0 & 0
	\end{bmatrix} \, ,
	&&
	c^{(3)} = \begin{bmatrix}
		-2 & 0 & 0 & 0 & 0 & 2 \\
		-2 & 0 & 0 & 0 & 2 & 0 \\
		-2 & 0 & 0 & 2 & 0 & 0 \\
		-2 & 1 & 1 & 0 & 0 & 0 \\
		-1 & 0 & 1 & 1 & 1 & 1 \\
		0 & 0 & 0 & 0 & 0 & 0
	\end{bmatrix} \, .
\end{align}

The classical bounds are $C= 3, 3$ and $6$, respectively and we know from above that they cannot be violated using qubit messages and qubit entanglement.  Using the alternating convex search method of Ref.~\cite{Tavakoli2021}, we find strategies using qubit messages and higher-dimensional entanglement that violate these facets. In Table \ref{tab:bounds6416}, we provide the quantum (lower) bounds up to $4\times 4$ entanglement with a qubit message.

\begin{table}[h!]
	\centering
	\begin{tabular}{|c|c|c|c|c|}
		\hline
		facet & $C$ & $Q_2^{(2\times 2)}$ & $Q_2^{(3\times 3)}$ & $Q_2^{(4\times 4)}$\\
		\hline
		1  & 3 & 3 & 3 & 3\\
		2  & 3 & 3 & 3 & 3.30\\
		3  & 6 & 6 & 6.051 & 6.31\\
		$I_{d+1,d,d+2}^{(1)}$  & 6 & 6 & 6.050 & 6.30 \\
		$I_{d+1,d,d+2}^{(2)}$  & 3 & 3 & 3 & 3\\
		\hline
	\end{tabular}
	\caption{Quantum (lower) bounds obtained for each facet class in the (6,4,1,6) scenario. $Q_2^{d\times d}$ is the bound for qubit communication assisted by $d\times d$ entanglement.}
	\label{tab:bounds6416}
\end{table}

For the experimental protocol, we looked for the best protocols that allow us to recycle the measurement setup of the task in the main text and are based on two copies of the singlet. We also restrict the minimal number of operators in the Kraus representation of Alice's channels to be at most two, making possible a unitary implementation that requires no additional ancillary qubits \cite{nielsen_chuang_2010}. 
This global unitary can then be decomposed into the two-qubit gates into a universal set of single-qubit gates and CNOT gates using the quantum Shannon decomposition \cite{Shende2006}. To this end, we used the Mathematica package \emph{UniversalQCompiler} \cite{Raban2019}. The resulting circuits contain 3 CNOT gates for each encoding operation and up to 12 elementary single-qubit gates, which makes it experimentally infeasible. At the cost of slightly reducing the maximal violation, we instead used gradient-based optimization algorithms (MATLAB) to look for an implementation based only on rotations in the X-Z plane and CNOT gates,
\begin{equation}
	R_{XZ}(\theta) = \begin{pmatrix}
		\cos(\theta) & -\sin(\theta)\\
		\sin(\theta) & \cos(\theta)
	\end{pmatrix}
\end{equation}
where $\theta\in [0,\pi]$. The best implementation we found achieves a score of $\mathcal{F} = 3.209$. It uses a sequence of gates for each unitary which consists of two CNOT gates sandwiched in between four single qubit rotations in the X-Z plane, two for each qubit. The unitaries corresponding to Alice's first input use three CNOTs in addition to the single qubit gates, while for the other inputs only two CNOTs are required.

The explicit decomposition is given in Fig. \ref{fig:ExpSetup2} with the same optical elements used in our experiment, while the angles for the single qubit gates are provided in Table \ref{table:FWimplementation}.

\begin{table}[h]
	\centering
	\caption{The angle settings of X-Z rotations to realize Alice's encoding unitary gate.}
	\begin{threeparttable}
		\begin{tabular}{c|c|c|c|c|c|c}
			\hline
			&\quad $U_1$ \quad&\quad $U_2$ \quad &\quad $U_3$ \quad &\quad $U_4$ \quad &\quad $U_5$ \quad &\quad $U_6$ \quad\\
			\hline
			\quad $R_{XZ}(\theta_1)$ \quad & $42.45\degree$ & $99.37\degree$ & $51.77\degree$ & $92.65\degree$ & $78.61\degree$ & $39.48\degree$ \\
			\hline
			\quad $R_{XZ}(\theta_2)$ \quad & $51.20\degree$ & $88.46\degree$ & $94.97\degree$ & $88.18\degree$ & $91.71\degree$ & $88.25\degree$ \\
			\hline
			\hline
			\quad $R_{XZ}(\theta_3)$ \quad & $14.79\degree$ & $45.04\degree$ & $67.34\degree$ & $90.05\degree$ & $157.48\degree$ & $0.27\degree$ \\
			\hline
			\quad $R_{XZ}(\theta_4)$ \quad & $84.71\degree$ & $89.66\degree$ & $86.84\degree$ & $43.33\degree$ & $179.65\degree$ & $87.23\degree$ \\
			\hline
			\hline
			\quad $R_{XZ}(\theta_5)$ \quad & $57.79\degree$ & $65.72\degree$ & $82.55\degree$ & $83.99\degree$ & $86.81\degree$ & $111.08\degree$ \\
			\hline
			\quad $R_{XZ}(\theta_6)$ \quad & $17.54\degree$  & $179.56$ & $173.3\degree$ & $48.55\degree$ & $92.67\degree$ & $0.03\degree$ \\
			\hline
			\hline
			\quad $R_{XZ}(\theta_7)$ \quad & $116.78\degree$ &  &  &  &  &  \\
			\hline
			\quad $R_{XZ}(\theta_8)$ \quad & $94.18\degree$ &  &  &  &  &  \\
			\hline
		\end{tabular}
		\begin{tablenotes}
			\item[*] For $U_4=\openone \otimes Z$, the HWP angle settings in the table can only realize $\openone \otimes \openone$. The required $Z$ gate is achieved by properly setting the SLM.
		\end{tablenotes}
	\end{threeparttable}
	\label{table:FWimplementation}
\end{table}

\bigskip
\begin{figure*}[tbph]
	\begin{center}
		\includegraphics [width=0.5\columnwidth]{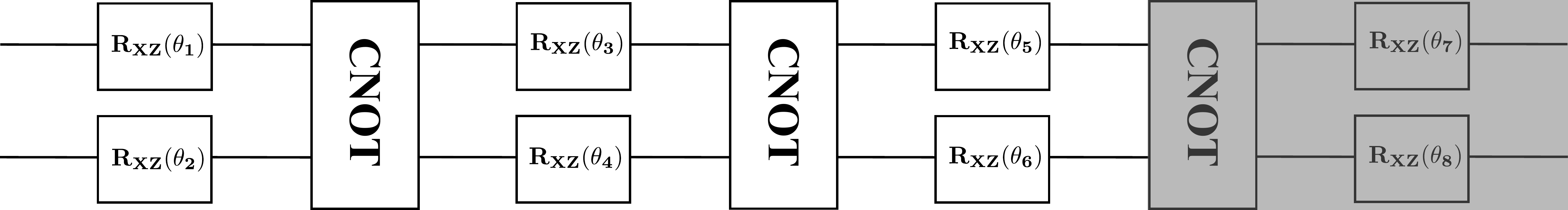}
	\end{center}
	\caption{\emph{Proposed circuit for the second facet inequality.} In our proposal, each channel consists of a circuit as depicted above, consisting only of CNOT operations and rotations in the X-Z plane. The unitary for the first channel, for $x=1$, consists of three CNOTs and eight rotations. The other channels only use two CNOTs and six rotations, (grey-shaded excluded).}
	\label{fig:FWCircuit}
\end{figure*}

\bigskip
\begin{figure*}[tbph]
	\begin{center}
		\includegraphics [width=0.95\columnwidth]{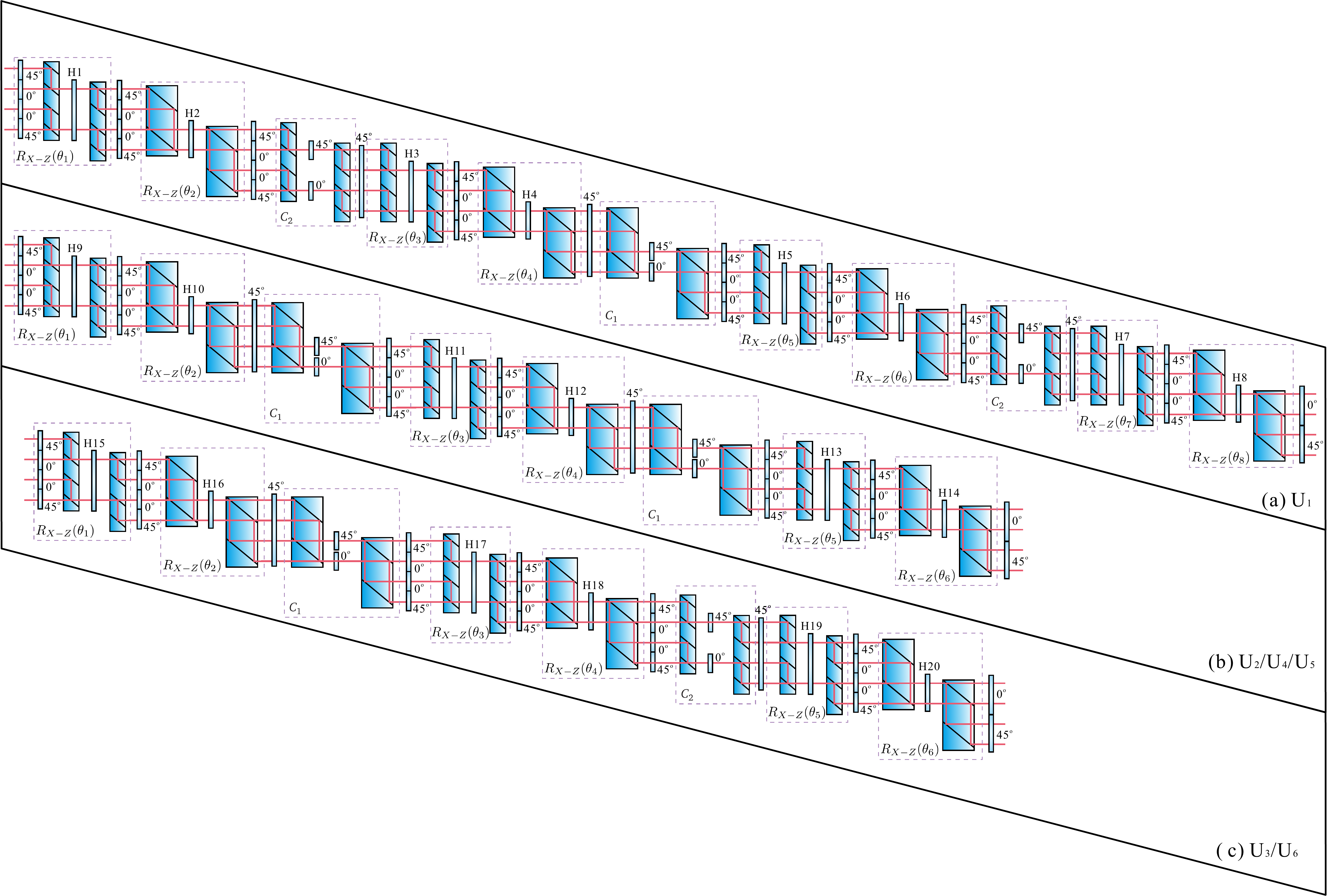}
	\end{center}
	\caption{\emph{Proposed setup for the second facet inequality.} $U_1$ (a), $U_{2, 4, 5}$ (b), and $U_{3, 6}$ (c) can be achieved by properly setting the angles of the HWPs (H1-H20). The sub-modules named $R_{XZ}(\theta_i)$ are used to realize rotations in the X-Z plane acting on the first (odd $i$) or the second qubit (even $i$). The sub-modules named $C_j$ are CNOT gates with the control on the $j$th qubit.}
	\label{fig:ExpSetup2}
\end{figure*}

\end{document}